\documentclass[a4paper,12pt]{article}
\pdfoutput=1 

\usepackage{jcappub} 
\usepackage[symbol]{footmisc}
\usepackage{changes}
\usepackage{amsmath,amssymb}
\usepackage{graphicx}
\usepackage{adjustbox}
\usepackage{multirow}
\usepackage[normalem]{ulem}
\usepackage{bm}        
\newcommand{\fpbh}{f_{\mathrm{PBH}}}

\usepackage{graphicx}   
\usepackage{tabulary}
\usepackage{color}

\newcommand{\nc}{\newcommand*}

\nc{\al}{\alpha}
\nc{\s}{\sigma}
\nc{\dt}{\delta}
\nc{\Dt}{\Delta}
\nc{\Ld}{\Lambda}
\nc{\p}{\partial}
\nc{\om}{\omega}
\nc{\Om}{\Omega}
\nc{\rd}{\mathrm{d}}
\nc{\Od}[1]{\mathcal{O}(#1)} 
\nc{\kp}{\kappa}
\nc{\fyr}{f_{\mathrm{yr}}}
\def\({\left(}
\def\){\right)}
\def\[{\left[}
\def\]{\right]}
\def\e{\begin{equation}}
\def\q{\end{equation}}
\def\m{\begin{eqnarray}}
\def\n{\end{eqnarray}}

\nc{\Eq}[1]{Eq.~\eqref{#1}}     
\nc{\Fig}[1]{Fig.~\ref{#1}}     
\nc{\Table}[1]{Table~\ref{#1}}  
\nc{\Sec}[1]{Sec.~\ref{#1}}     

\nc{\Msun}{M_\odot}             
\nc{\fpbhn}{f_{\mathrm{pbh0}}}    
\nc{\mR}{\mathcal{R}} 
\nc{\seq}{\sigma_{\mathrm{eq}}}
\nc{\ogw}{\Omega_{\mathrm{GW}}}
\nc{\gpcyr}{\mathrm{Gpc}^{-3}\,\mathrm{yr}^{-1}}
\nc{\lvc}{LIGO/Virgo} 
\nc{\SNR}{\mathrm{SNR}} 
\nc{\mmin}{{m_{\mathrm{min}}}}
\nc{\mmax}{{m_{\mathrm{max}}}}
\nc{\Mmin}{{M_{\mathrm{min}}}}
\nc{\fmin}{{f_{\mathrm{min}}}}
\nc{\VT}{\mathrm{VT}}
\nc{\rhoGW}{\rho_{\mathrm{GW}}}
\nc{\vth}{\vec{\theta}}
\nc{\vd}{\vec{d}}
\nc{\vla}{\vec{\lambda}}
\nc{\Nobs}{N_{\mathrm{obs}}}
\nc{\av}[1]{\langle #1 \rangle} 
\nc{\km}{\mathrm{km}}
\nc{\Mpc}{\mathrm{Mpc}}
\nc{\Tobs}{T_{\mathrm{obs}}}
\nc{\Ntemp}{N_{\mathrm{temp}}}
\nc{\ie}{\textit{i.e.}}
\nc{\addref}{[\textcolor{red}{add ref}] } 
\nc{\eg}{\textit{e.g.~}}
\nc{\app}{\approx}
\nc{\hf}{\frac{1}{2}}
\nc{\discuss}{\textcolor{red}{Add discussion here!}}

\nc{\mpbh}{m_{\rm{pbh}}}
\nc{\cR}{\mathcal{R}}
\nc{\mU}{{\mathcal{U}}}
\nc{\Mc}{{M_\mathrm{c}}}
\nc{\Mf}{{M_\mathrm{f}}}
\nc{\red}[1]{\textcolor{red}{#1}}
\nc{\yellow}[1]{\textcolor{yellow}{#1}}
\nc{\green}[1]{\textcolor{green}{#1}}
\nc{\blue}[1]{\textcolor{blue}{#1}}
\nc{\fnl}{F_{\mathrm{NL}}}
\nc{\gnl}{G_{\mathrm{NL}}}
\nc{\MG}{\mathcal{M}_{\mathrm{G}}}
\nc{\MNG}{\mathcal{M}_{\mathrm{NG}}}

\begin{document}
	
\title{Simultaneously probing the sound speed and equation of state of the early Universe with pulsar timing arrays} 
\author{Lang~Liu,$^{a,b}$}
\author{You~Wu,\note{Corresponding author.}$^{c,d,*}$}
\author{and Zu-Cheng~Chen$^{d,e,a,b,*}$}

\affiliation{$^a$Department of Astronomy, Beijing Normal University, Beijing 100875, China}
\affiliation{$^b$Advanced Institute of Natural Sciences, Beijing Normal University, Zhuhai 519087, China}
\affiliation{$^c$College of Mathematics and Physics, Hunan University of Arts and Science, Changde, 415000, China}
\affiliation{$^d$Department of Physics and Synergetic Innovation Center for Quantum Effects and Applications, Hunan Normal University, Changsha, Hunan 410081, China}
\affiliation{$^e$Institute of Interdisciplinary Studies, Hunan Normal University, Changsha, Hunan 410081, China}
\emailAdd{liulang@bnu.edu.cn}	
\emailAdd{youwuphy@gmail.com}	
\emailAdd{zuchengchen@gmail.com}
	
\abstract{
Recently, several major pulsar timing array (PTA) collaborations have assembled strong evidence for the existence of a gravitational-wave background at frequencies around the nanohertz regime. Assuming that the PTA signal is attributed to scalar-induced gravitational waves, we jointly employ the PTA data from the NANOGrav 15-year data set, PPTA DR3, and EPTA DR2 to probe the conditions of the early Universe.
Specifically, we explore the equation of state parameter ($w$), the reheating temperature ($T_\mathrm{rh}$), and the sound speed ($c_s$), finding $w = 0.59^{+0.36}_{-0.40}$ (median + $90\%$ credible interval), and $T_\mathrm{rh}\lesssim 0.2\,\mathrm{GeV}$ at the $95\%$ credible interval for a lognormal power spectrum of the curvature perturbation. 
Furthermore, we compute Bayes factors to compare different models against the power-law spectrum model, effectively excluding the pressure-less fluid domination model. Our study underscores the significance of scalar-induced gravitational waves as a powerful tool to explore the nature of the early Universe.
}

\maketitle
\section{Introduction}

The groundbreaking detection of gravitational waves (GWs) originating from compact binary mergers observed by ground-based detectors~\cite{LIGOScientific:2018mvr,LIGOScientific:2020ibl,LIGOScientific:2021djp}, has significantly advanced our ability to test gravity theories within the realm of strong gravitational fields~\cite{LIGOScientific:2019fpa,LIGOScientific:2020tif,LIGOScientific:2021sio}. Furthermore, it has enabled us to explore the characteristics of the broader population of GW sources~\cite{LIGOScientific:2018jsj,Chen:2018rzo,Chen:2019irf,LIGOScientific:2020kqk,Chen:2021nxo,KAGRA:2021duu,Chen:2022fda,Liu:2022iuf,Zheng:2022wqo,You:2023ouk}. On the other hand, the detection of a stochastic gravitational wave background (SGWB) remains an ongoing pursuit. The pulsar timing array (PTA) serves as an indispensable tool for probing the SGWB in the nanohertz frequency range, offering a valuable window into the detection of GWs that originated from the early Universe. 
Recently, several PTA collaborations, including the North American Nanohertz Observatory for Gravitational Waves (NANOGrav)~\cite{NANOGrav:2023gor,NANOGrav:2023hde}, the Parkers PTA (PPTA)~\cite{Zic:2023gta,Reardon:2023gzh}, the European PTA (EPTA) along with the Indian PTA (InPTA)~\cite{EPTA:2023sfo,Antoniadis:2023ott}, and the Chinese PTA (CPTA)~\cite{Xu:2023wog}, have independently presented compelling evidence for the spatial correlations consistent with the Hellings-Downs~\cite{Hellings:1983fr} pattern in their latest data sets. These correlations align with the predicted characteristics of an SGWB within the framework of general relativity. While there exists various potential sources within the PTA window~\cite{Li:2019vlb,Vagnozzi:2020gtf,Chen:2021wdo,Wu:2021kmd,Chen:2021ncc,Sakharov:2021dim,Benetti:2021uea,Chen:2022azo,Ashoorioon:2022raz,PPTA:2022eul,Wu:2023pbt,IPTA:2023ero,Wu:2023dnp,Dandoy:2023jot,Madge:2023cak,Chen:2023zkb,Yi:2023npi,Wu:2023rib,Bi:2023ewq,Chen:2023uiz}, the nature of this signal, whether it originates from astrophysical or cosmological phenomena, remains the subject of intensive investigation~\cite{NANOGrav:2023hvm,Antoniadis:2023xlr,King:2023cgv,Niu:2023bsr,Bi:2023tib, Liu:2023pau,Vagnozzi:2023lwo,Fu:2023aab,Han:2023olf,Li:2023yaj,Franciolini:2023wjm,Shen:2023pan,Kitajima:2023cek,Franciolini:2023pbf,Addazi:2023jvg,Cai:2023dls,Inomata:2023zup,Murai:2023gkv,Li:2023bxy,Anchordoqui:2023tln,Liu:2023ymk,Abe:2023yrw,Ghosh:2023aum,Figueroa:2023zhu,Yi:2023mbm,Wu:2023hsa,Bian:2023dnv,Li:2023tdx,Geller:2023shn,You:2023rmn,Antusch:2023zjk,Ye:2023xyr,HosseiniMansoori:2023mqh,Jin:2023wri,Zhang:2023nrs,ValbusaDallArmi:2023nqn,DeLuca:2023tun,Choudhury:2023kam,Gorji:2023sil,Das:2023nmm,Yi:2023tdk,Ellis:2023oxs,He:2023ado,Balaji:2023ehk,Kawasaki:2023rfx,Cannizzaro:2023mgc,King:2023ayw,Maji:2023fhv,Bhaumik:2023wmw,Zhu:2023lbf,Basilakos:2023xof,Huang:2023chx,Jiang:2023gfe,DiBari:2023upq,Aghaie:2023lan,Garcia-Saenz:2023zue,InternationalPulsarTimingArray:2023mzf,Harigaya:2023pmw,Altavista:2023zhw,Lozanov:2023rcd,Choudhury:2023fwk,Cang:2023ysz,Chen:2023bms,Chen:2024fir,Fei:2023iel}.

One plausible explanation for the observed signal is the scalar-induced gravitational waves (SIGWs), generated by primordial curvature perturbations at small scales~\cite{Ananda:2006af,Baumann:2007zm,Garcia-Bellido:2016dkw,Inomata:2016rbd,Garcia-Bellido:2017aan,Kohri:2018awv,Cai:2018dig,Lu:2019sti,Yuan:2019wwo,Chen:2019xse,Xu:2019bdp,Yuan:2019udt,Cai:2019cdl,Yuan:2019fwv,Yi:2020kmq,Yi:2020cut,Liu:2020oqe,Gao:2020tsa,Yuan:2020iwf,Yuan:2021qgz,Yi:2021lxc,Yi:2022anu,Yi:2022ymw,Yuan:2023ofl,Meng:2022ixx}. This explanation is supported by the NANOGrav data, which favor the SIGWs scenario over the supermassive black hole binaries (SMBHBs) scenario through Bayesian analysis~\cite{NANOGrav:2023hvm}. When primordial curvature perturbations reach significant magnitudes, they can generate a substantial SGWB through second-order effects resulting from the nonlinear coupling of perturbations. Additionally, the presence of large curvature perturbations can trigger the formation of primordial black holes (PBHs)~\cite{Zeldovich:1967lct,Hawking:1971ei,Carr:1974nx}.
PBHs have attracted considerable attention in recent years~\cite{Belotsky:2014kca,Carr:2016drx,Garcia-Bellido:2017mdw,Carr:2017jsz,Germani:2017bcs,Chen:2018rzo,Liu:2018ess,Chen:2018czv,Liu:2019rnx,Fu:2019ttf,Liu:2019lul,Cai:2019bmk,Chen:2019irf,Liu:2020cds,Fu:2020lob,Wu:2020drm,DeLuca:2020sae,Vaskonen:2020lbd,DeLuca:2020agl,Domenech:2020ers,Hutsi:2020sol,Chen:2021nxo,Kawai:2021edk,Braglia:2021wwa,Cai:2021wzd,Liu:2021jnw,Braglia:2022icu,Zheng:2022wqo,Chen:2022qvg,Liu:2022iuf,Chen:2022fda,Inomata:2022yte,Guo:2023hyp,Cai:2023uhc,Meng:2022low,Gu:2023mmd} (see also reviews~\cite{Sasaki:2018dmp,Carr:2020gox,Carr:2020xqk}), as they are potential candidates for dark matter~\cite{Sasaki:2018dmp,Carr:2020gox,Carr:2020xqk}. Moreover, PBHs offer a promising explanation for the observed binary black holes detected by LIGO-Virgo-KAGRA~\cite{Bird:2016dcv,Sasaki:2016jop}.

The infrared power-law behavior of SIGWs is influenced by the equation of state (EoS), $w$, of the early Universe when the corresponding wavelength of primordial curvature perturbations reenters the Hubble horizon~\cite{Domenech:2019quo,Domenech:2020kqm,Domenech:2021ztg}. Additionally, the sound speed $c_s$ can impact the amplitude and resonant peak of SIGWs~\cite{Domenech:2021ztg}.
Consequently, SIGWs provide an invaluable opportunity to directly probe the expansion history of the primordial dark Universe.
Multiple aspects of SIGWs offer avenues for exploring potential new physics, including the exploration of different equations of state for the Universe~\cite{Domenech:2020ssp,Domenech:2021wkk,Domenech:2019quo,Domenech:2020kqm,Liu:2023pau,Zhao:2023joc}, the investigation of different propagation speeds of fluctuations~\cite{Domenech:2021ztg,Balaji:2022dbi,Balaji:2023ehk}, and the examination of different initial conditions~\cite{Domenech:2021and}. 
In particular, Ref.~\cite{Balaji:2023ehk} suggests that if the recent PTA signal is attributed to SIGWs, the Universe can be characterized as a perfect fluid with $w=1/3$ and an arbitrary $c_s$. 
Meanwhile, Ref.~\cite{Liu:2023pau} and Ref.~\cite{Zhao:2023joc} investigate the EoS parameters of the early Universe using PTA data but assume different scenarios. Ref.~\cite{Liu:2023pau} assumes that the early Universe was dominated by a canonical scalar field with $c_s^2=1$, while Ref.~\cite{Zhao:2023joc} assumes an adiabatic perfect fluid with $c_s^2=w$. To gain a deeper understanding of the early Universe, it is crucial to simultaneously impose constraints on both the sound speed and EoS parameters.

In this paper, we investigate a scenario in which the Universe, prior to the end of reheating, exhibits an arbitrary EoS parameter and sound speed. This scenario can be realized by a scalar field undergoing oscillations in an arbitrary potential~\cite{Lucchin:1984yf,Balaji:2023ehk}. Under the assumption that the stochastic signal detected by PTAs originates from SIGWs, we conduct a comprehensive analysis by jointly employing PTA data from the NANOGrav 15-year data set, PPTA DR3, and EPTA DR2 to simultaneously impose constraints on both the sound speed and EoS of the early Universe. The rest of this paper is organized as follows. In Section~\ref{SIGW}, we provide an overview of the energy density of SIGWs generated in a Universe with general $w$ and $c_s$, along with the PBH production. 
In Section~\ref{Data}, we provide a detailed description of the methodology employed for data analysis and present the results obtained using the most recent PTA data sets. Additionally, we compute the Bayes factors comparing various models with the power-law spectrum model where $\Omega_{\mathrm{GW}} \propto A^2 f^{-\gamma}$. Finally, we summarize our findings and give some discussions in Section~\ref{Con}.

\section{Scalar-induced gravitational waves and primordial black holes} 
\label{SIGW}
\begin{figure}[tbp!]
	\centering
	\includegraphics[width=\textwidth]{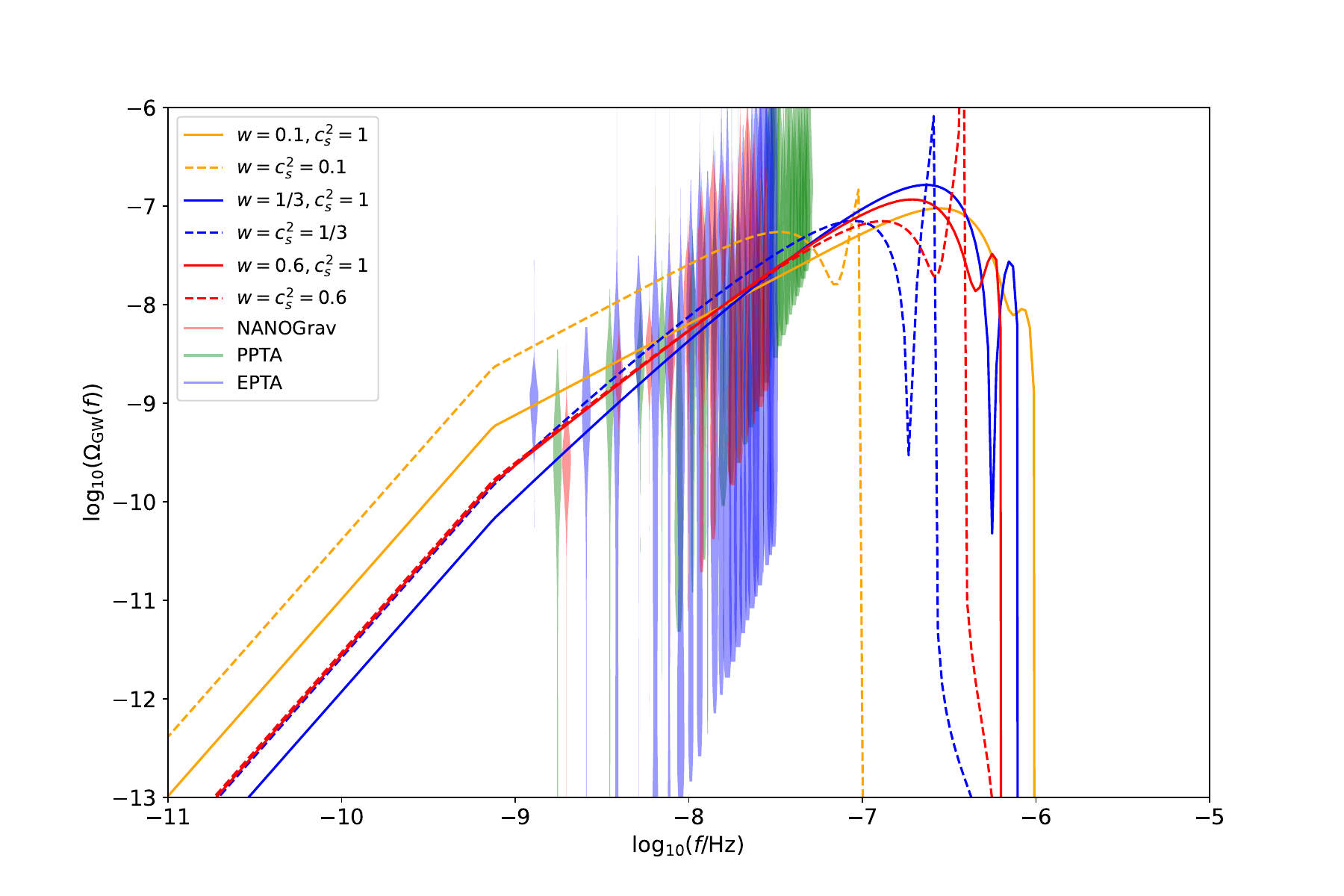}
	\caption{\label{ogw}The energy density spectrum of SIGW with different EoS parameter $w$ and sound speed $c_s$ as a function of GW frequency.}
\end{figure}

In order to generate a detectable SGWB, the scalar perturbations must undergo a significant amplification compared to the fluctuations observed in cosmic microwave background (CMB) experiments. The amplitude and shape of the SIGW spectrum are determined by the properties of primordial fluctuations, including the power spectrum and non-Gaussianities, as well as the composition of the Universe after inflation. These factors collectively influence the characteristics of the SIGW spectrum, shaping its amplitude and shape. In order to model the primordial spectrum, we adopt a log-normal peak description for the small-scale enhancement. Specifically, we express the spectrum as
\begin{equation}
\label{PR}
\mathcal{P}_{\mathcal{R}} (k) = \frac{A}{\sqrt{2\pi}\Delta} \exp\! \left( -\frac{\ln^{2}(k/k_{*})}{2\Delta^{2}} \right).
\end{equation}
Here, $A$ represents the amplitude of the spectrum, $k_*$ denotes the characteristic scale, and $\Delta$ corresponds to the width of the spectrum. In our analysis, we focus on the narrow peak spectrum, where $\Delta \lesssim 0.1$, which is supported by our results. As $\Delta$ approaches zero, the primordial power spectrum $\mathcal{P}_{\mathcal{R}}$ tends to a $\delta$-function form, $\mathcal{P}_{\mathcal{R}} = A k_* \delta(k-k_*)$. The effect of a finite width for SIGWs can be expressed as~\cite{Pi:2020otn}
\begin{equation}\label{eq:OmegaDelta}
\Omega_{\rm GW,0}h^2={\rm Erf}\!\left[\frac{1}{\Delta}\sinh^{-1}\frac{k}{2k_{\rm *}}\right]\Omega^\delta_{\rm GW,0}h^2.
\end{equation}
Here, $\Omega_{\rm GW,0}h^2$ represents the present-day GW energy density fraction with finite width, ${\rm Erf}$ denotes the error function, $\Omega^\delta_{\rm GW,0}h^2$ corresponds to the GW spectrum induced by a $\delta$-function peak. In the following, we adopt an instantaneous reheating scenario assuming that the Universe undergoes reheating immediately when the mode with comoving wave number $k_\mathrm{rh}$ reenters the horizon. When considering Gaussian fluctuations, there exists a general integral formula for the SIGW spectrum in the case of a constant EoS parameter $w$ and sound speed $c_s$. The present-day fraction of GW energy density reads
\begin{equation}\label{ogw0}
\begin{split}
\Omega_{\mathrm{GW}, 0} h^2 \approx & 1.62 \times 10^{-5}\left(\frac{\Omega_{r, 0} h^2}{4.18 \times 10^{-5}}\right)   \left(\frac{g_{*r}(T_{\mathrm{rh}})}{106.75}\right)\left(\frac{g_{* s}(T_{\mathrm{rh}})}{106.75}\right)^{-\frac{4}{3}} \Omega_{\mathrm{GW}, \mathrm{rh}},    
\end{split}
\end{equation}
where $\Omega_{r,0}h^2$ denotes the present-day radiation fraction, $g_{*r}$ and $g_{*s}$ are the effective energy and entropy degrees of freedom, respectively. We refer to Ref.~\cite{Saikawa:2018rcs} for precise numerical fits. The subscript ``rh" in our notation refers to reheating and specifically denotes the time of the instantaneous transition to the standard radiation-dominated era. The SIGW spectrum for scales $k \gtrsim k_{\rm rh}$, as given by Eq.~\eqref{ogw0}, can be expressed as~\cite{Domenech:2021ztg}
\begin{equation}
\Omega_{\mathrm{GW},\mathrm{rh}}=\left(\frac{k}{k_{\mathrm{rh}}}\right)^{-2 b} \int_0^{\infty} d v \int_{|1-v|}^{1+v} d u\, \mathcal{T}(u,v,b,c_s)\, \mathcal{P}_{\mathcal{R}}(k u)\, \mathcal{P}_{\mathcal{R}}(k v),
\end{equation}
where $b\equiv (1-3w)/(1+3w)$ and the transfer function is defined as 
\begin{align}\label{eq:kernelaveragefinal}
{\cal T}(u,v,b,c_s)=& {\cal N}(b,c_s)\left(\frac{4v^2-(1-u^2+v^2)^2}{4u^2v^2}\right)^2{|1-y^2|^{b}}\nonumber\\&
\times\Bigg\{\left(\mathsf{P}^{-b}_{b}(y)+\frac{b+2}{b+1}\mathsf{P}^{-b}_{b+2}(y)\right)^2\Theta(c_s(u+v)-1)\nonumber\\&
+\frac{4}{\pi^2}\left(\mathsf{Q}^{-b}_{b}(y)+\frac{b+2}{b+1}\mathsf{Q}^{-b}_{b+2}(y)\right)^2\Theta(c_s(u+v)-1)\nonumber\\&
+\frac{4}{\pi^2}\left({\cal Q}^{-b}_{b}(-y)+2\frac{b+2}{b+1}\mathsf{\cal Q}^{-b}_{b+2}(-y)\right)^2\Theta(1-c_s(u+v))\Bigg\},
\end{align}
where $\Theta$ is the Heaviside theta function, $y\equiv 1-[1-c_s^2(u-v)^2]/(2c_s^2 uv)$, $\mathsf{P}^{\mu}_{\nu}(x)$ and $\mathsf{Q}^{\mu}_{\nu}(x)$ are Ferrer's functions,  ${\cal Q}^{\mu}_{\nu}(x)$ is Olver's function, and the numerical coefficient is
\begin{align}
{\cal N}(b,c_s)\equiv \frac{4^{2b}}{3c_s^4}\Gamma(b+\tfrac{3}{2})^4\left(\frac{b+2}{2b+3}\right)^2\left(1+b\right)^{-2(1+b)}\,.
\end{align}
Here, $\Gamma$ is the Gamma function.
The transfer function given by \Eq{eq:kernelaveragefinal} is expressed in terms of associated Legendre functions. These functions can be expressed in terms of hypergeometric functions as
\begin{align}
\mathsf{P}^{\mu}_{\nu}(x)=\left(\frac{1+x}{1-x}\right)^{\mu/2}%
\mathbf{F}(\nu+1,-\nu;1-\mu;\tfrac{1}{2}-\tfrac{1}{2}x)\,,
\end{align}
\begin{align}
\mathsf{Q}^{\mu}_{\nu}(x)=&\frac{\pi}{2\sin\left(\mu\pi\right)}\Bigg%
(\cos\left(\mu\pi\right)\left(\frac{1+x}{1-x}\right)^{\mu/2}\mathbf{F}(%
\nu+1,-\nu;1-\mu;\tfrac{1}{2}-\tfrac{1}{2}x)\nonumber\\&\qquad\qquad\qquad-\frac{\Gamma(\nu+\mu+1%
)}{\Gamma(\nu-\mu+1)}\left(\frac{1-x}{1+x}\right)^{\mu/2}%
\mathbf{F}(\nu+1,-\nu;1+\mu;\tfrac{1}{2}-\tfrac{1}{2}x)\Bigg)\,,
\end{align}
and
\begin{align}
{\cal Q}^{\mu}_{\nu}(x)=&%
\frac{\pi}{2\sin\left(\mu\pi\right)\Gamma(\nu+\mu+1)}\Bigg(\left(\frac{x+1}{x-1}\right)^{\mu/2}\mathbf{F}(\nu+1,-\nu;1-\mu;\tfrac{1}{2}-\tfrac{1}{2}x)\nonumber\\&\qquad\qquad\qquad-\frac{\Gamma(\nu+\mu+1)}{%
\Gamma(\nu-\mu+1)}\left(\frac{x-1}{x+1}\right)^{\mu/2}\mathbf{F}(\nu+1,-\nu;1+\mu;%
\tfrac{1}{2}-\tfrac{1}{2}x)\Bigg)\,.
\end{align}
It is worth noting that the functions $\mathsf{P}^{\mu}_{\nu}(x)$ and $\mathsf{Q}^{\mu}_{\nu}(x)$ are valid for $|x|<1$. Additionally, the function ${\cal Q}^{\mu}_{\nu}(x)$ is real for $|x|>1$. In the previous definitions, we employed the concise notation that
\begin{align}
\mathbf{F}(a,b;c;x)=\frac{1}{\Gamma(c)}{F}(a,b;c;x),
\end{align}
where ${F}(a,b;c;x)$ represents Gauss's hypergeometric function.
Since the SIGW spectrum, $\Omega_{\mathrm{GW},\mathrm{rh}}$, is proportional to $(k/k_{\rm rh})^2$~\cite{Domenech:2020kqm,Domenech:2021ztg,Liu:2023pau}, the amplitude of the SIGW spectrum is suppressed by a factor of $(k_{\rm rh}/k_*)^2$. We illustrate the energy density spectrum of SIGWs as a function of $w$ and $c_s$ in \Fig{ogw}. Now, let's derive the relation between the frequency $f$ and the temperature $T$. We can estimate that each mode with a wavenumber $k$ crosses the horizon at temperature $T$ using the following approximation
\begin{equation}
\label{k-T}
    k\! \simeq \frac{\!1.5\!\times\!  10^7}{\mathrm{Mpc}}
    \left({g_{*r}(T)} \over {106.75} \right)^\frac{1}{2}\!\!
    \left({g_{*s}(T)} \over 106.75\right)^{-\frac{1}{3}}\!\!
    \left (\frac{T}{\rm GeV}\right),
\end{equation}
with the corresponding GW frequency $f$ at the current epoch for each mode $k$,
\begin{equation}
\label{k-f}
f = \frac{k}{2 \pi} \simeq 1.6\, \mathrm{nHz}
\left(\frac{k}{10^6\,\mathrm{Mpc}^{-1}}\right).
\end{equation}
Therefore, combining \Eq{k-T} and \Eq{k-f}, we can establish the relation between the frequency $f$ and the temperature $T$ as
\begin{equation}
f \simeq 24\, \mathrm{nHz} \left(\frac{g_{*r}(T)}{106.75}\right)^{\frac{1}{2}}
\left(\frac{g_{*s}(T)}{106.75}\right)^{-\frac{1}{3}}
\left(\frac{T}{\mathrm{GeV}}\right).
\end{equation}
It is worth noting that the lower bound on the reheating temperature for BBN is $T_{\mathrm{rh}} \geq 4\,\mathrm{MeV}$~\cite{Kawasaki:1999na,Kawasaki:2000en,Hannestad:2004px,Hasegawa:2019jsa}. Consequently, the constraints on the reheating frequency $f_{\mathrm{rh}}$ for the reheating mode $k_{\mathrm{rh}}$ are $f_{\mathrm{rh}} \gtrsim 0.1\,\mathrm{nHz}$, which is smaller than the sensitivity frequencies of PTAs. 

On the other hand, if the primordial curvature perturbations attain a sufficient magnitude, they can give rise to the formation of PBHs~\cite{Zeldovich:1967lct,Hawking:1971ei,Carr:1974nx,Meszaros:1974tb,Carr:1975qj,Musco:2004ak,Musco:2008hv,Musco:2012au,Harada:2013epa,Escriva:2020tak}. The abundance of PBHs can be described using the Press-Schechter formalism, which postulates that collapse occurs when the density contrast $\delta=\delta\rho/\rho$ exceeds a critical value $\delta_c$. It can be expressed mathematically as
\begin{align}
\label{eq:press}
\beta=\frac{\gamma}{\sqrt{2\pi\sigma^2}}\int_{\delta_c}^\infty e^{-\delta^2/(2\sigma^2)}d\delta=\frac{\gamma}{2}\text{erfc}\left(\frac{\delta_c(w)}{\sqrt2\sigma(M)}\right),
\end{align}
where $\gamma\approx0.2$ represents the efficiency factor of the collapse, which denotes the fraction of matter within the Hubble horizon that undergoes collapse to form PBHs. The integral represents the probability of collapse for values of $\delta$ greater than $\delta_c$, and it can be alternatively expressed using the complementary error function. In the given expression, we assume a Gaussian probability distribution for the fluctuations. The smoothed variance of the density contrast $\sigma$ can be related to the dimensionless primordial spectrum of curvature perturbations $\mathcal{P}_\mathcal{R}$ through the gradient expansion ~\cite{Harada:2015yda}
\begin{equation}
\label{sigma}
\sigma^{2} =\left(\frac{2+2w}{5+3w}\right)^{2} \int_{0}^{\infty} \frac{\mathrm{d}q}{q}\, \tilde{W}^{2}(\frac{q}{k})\left(\frac{q}{k}\right)^{4} T^{2}(\frac{q}{k})\, \mathcal{P}_{\mathcal{R}}(q).
\end{equation}
Here, $T(q/k)=3(\sin l - l \cos l)/l^{3}$ represents the transfer function with $l = q/k/\sqrt{3}$, and $\tilde{W}(q/k)=\exp(-q^{2}/k^{2}/2)$ denotes the window function, which is taken as a Gaussian. The dependence of the critical value $\delta_c$ in Eq.\,(\ref{eq:press}) on the EoS parameter $w$ and sound speed $c_s$ can be roughly estimated as~\cite{Domenech:2020ers,Balaji:2023ehk}
\begin{align}\label{dcb}
\delta_c \simeq \frac{3(1+w)}{5+3w} c_s^2.
\end{align}
Note that this formula is not applicable for the limit $c_s\rightarrow 0$ due to the significant influence of non-sphericity and angular momentum of the collapsing cloud~\cite{Harada:2016mhb,Harada:2017fjm}. In a Universe dominated by a general $w$ component, the masses of PBHs are related to the comoving scale by the following equation:
\begin{align}\label{mpbh}
\frac{M_{\rm PBH}}{M_\odot}&\approx  0.01\frac{\gamma}{0.2}\!\left(\frac{k_{\rm rh}}{k_*}\right)^{\!\!\!\frac{3(1+w)}{1+3w}}\!\!\! \left(\frac{106.75}{g_{*r}(T_{\rm rh})}\right)^{\!\!\frac12}\!\!\!\left(\frac{{\rm GeV}}{T_{\rm rh}}\right)^{\!\!\! 2}.
\end{align}
In the case of a sharply peaked primordial scalar power spectrum, resulting in a monochromatic mass function for the PBHs, the abundance of PBHs, expressed as the PBH energy fraction with respect to cold dark matter, can be described by
\begin{equation}\label{fpbh}
\begin{split}
f_{\rm PBH}\equiv\frac{\Omega_\text{PBH}}{\Omega_\text{CDM}} \approx & 1.5\times 10^{13}\beta\left(\frac{k_*}{k_{\rm rh}}\right)^{\frac{6w}{1+3w}}\left(\frac{T_{\rm rh}}{{\rm GeV}}\right)
\left(\frac{g_{*s}(T_{\rm rh})}{106.75}\right)^{-1}\left(\frac{g_{*r}(T_{\rm rh})}{106.75}\right).   
\end{split}
\end{equation}
Here, we remark that recent studies have explored alternative approaches, such as the peak theory formalism, to calculate the abundance of PBHs instead of the traditional Press-Schechter formalism, see e.g.\,\cite{Riccardi:2021rlf}.

\section{Data analyses and results}
\label{Data}

\begin{table}
    \centering
	\begin{tabular}{c|ccccccc}
		\hline\hline
		Parameter & $\log_{10} (f_*/\mathrm{Hz})$ & $\log_{10} \Delta$ & $\log_{10} A$ & $\log_{10} (T_{\mathrm{rh}}/\mathrm{Gev})$ & $w$ & $c_s$\\[1pt]
		\hline
		 Prior& $\mU(-10, -2)$ & $\mU(-6, -1)$ & $\mU(-5, 1)$ & $\mU(\log_{10} 0.004, 0.6)$ &  $\mU(-1/3, 1.2)$  &  $\mU(0, 1)$\\[1pt]
		Result  & $-5.52^{+2.00}_{-1.15}$ & $\lesssim -2.60$ & $-2.23^{+2.07}_{-1.30}$ & $\lesssim -0.7$ & $0.59^{+0.36}_{-0.40}$& $\gtrsim 0.09$\\[1pt]
  \hline
	\end{tabular}
	\caption{\label{tab:priors}Prior distributions and results for the model parameters. Here, $\mU$ represents the uniform distribution. The results are quoted in median value and $90\%$ equal-tail credible interval for each parameter.}
\end{table}

\begin{table}
    \centering
	\begin{tabular}{c|c|c|c|c|c|c|c}
		\hline\hline
		\multirow{2}{*}{Model} & $M_0$ &$M_1$ & $M_2$ & $M_3$ & $M_4$ & $M_5$ & $M_6$ \\
  & $c_s^2=w=1/3$& $c_s$, $w$ & $c_s^2 =0.47$, $w=0.59$ & $c_s^2 =1$, $w$ & $c_s^2 = w$ & $c_s$, $w=1/3$ & $c_s$, $w=0$ \\[1pt]
		\hline
		 BF & $0.81$& $0.85$ & $2.87$ & $0.96$ & $1.20$ & $0.87$ & $0.0055$  \\[1pt]
  \hline
	\end{tabular}
	\caption{\label{tab:BF}Bayes factors comparing different models with the reference model $M_\mathrm{ref}$ that characterize a power-law SGWB spectrum such that $\Omega_{\mathrm{GW}} \propto A^2 f^{-\gamma}$.}
\end{table}

In this work, we jointly use the NANOGrav 15-year data set~\cite{NANOGrav:2023hde}, PPTA DR3~\cite{Zic:2023gta}, and EPTA DR2~\cite{EPTA:2023sfo} to estimate the model parameters. Specifically, we utilize the free spectrum amplitudes obtained by each PTA when considering spatial correlations of the Hellings-Downs pattern. The sensitivity of a PTA's observations begins at a frequency of $1/T_{\mathrm{obs}}$, where $T_{\mathrm{obs}}$ is the observational time span. 
NANOGrav, PPTA, and EPTA employ $14$~\cite{NANOGrav:2023gor}, $28$~\cite{Reardon:2023gzh}, and $24$~\cite{Antoniadis:2023rey} frequencies, respectively, in their search for the SGWB signal.
When combining the data from these PTAs, we work with a total of $66$ frequency components in the free spectrum, spanning from $1.28$\,nHz to $49.1$\,nHz. In \Fig{ogw}, we illustrate the data employed in our analyses and depict the energy density of SIGWs for various values of $w$ and $c_s$. 

We commence our analysis with the posterior data of time delay $d(f)$, which is provided by each PTA. The power spectrum $S(f)$ is related to the time delay through 
\begin{equation}
S(f) = d(f)^2\, T_{\mathrm{obs}}.
\end{equation}
With this time delay data, we can calculate the energy density of the free spectrum by
\begin{equation}
\hat{\Omega}_{\mathrm{GW}}(f)=\frac{2 \pi^2}{3 H_0^2} f^2 h_c^2(f) = \frac{8\pi^4}{H_0^2} T_{\mathrm{obs}} f^5 d^2(f),
\end{equation}
where $H_0$ stands for the Hubble constant.
The characteristic strain, $h_c(f)$, is defined as
\begin{equation}
h_c^2(f)=12 \pi^2 f^3 S(f).
\end{equation}
For each observed frequency $f_i$, using the obtained posteriors of $\hat{\Omega}_{\mathrm{GW}}(f_i)$ as described above, we can estimate the corresponding kernel density, $\mathcal{L}_i$. Therefore, the total log-likelihood is the sum of individual likelihoods given by~\cite{Liu:2023ymk,Wu:2023hsa,Jin:2023wri,Liu:2023pau}
\begin{equation}
\ln \mathcal{L}(\Lambda) = \sum_{i=1}^{66} \ln \mathcal{L}_i(\Omega_{\mathrm{GW}}(f_i, \Lambda)),
\end{equation}
where $\Lambda\equiv \{A, \Delta, f_*, T_{\mathrm{rh}}, w, c_s\}$ denotes the set of six model parameters.
To explore the parameter space, we employ the \texttt{dynesty}~\cite{Speagle:2019ivv} sampler available in the \texttt{Bilby}~\cite{Ashton:2018jfp,Romero-Shaw:2020owr} package. A summary of the priors and results for the model parameters is presented in~\Table{tab:priors}.

\begin{figure}[tbp!]
	\centering
 \includegraphics[width=\textwidth]{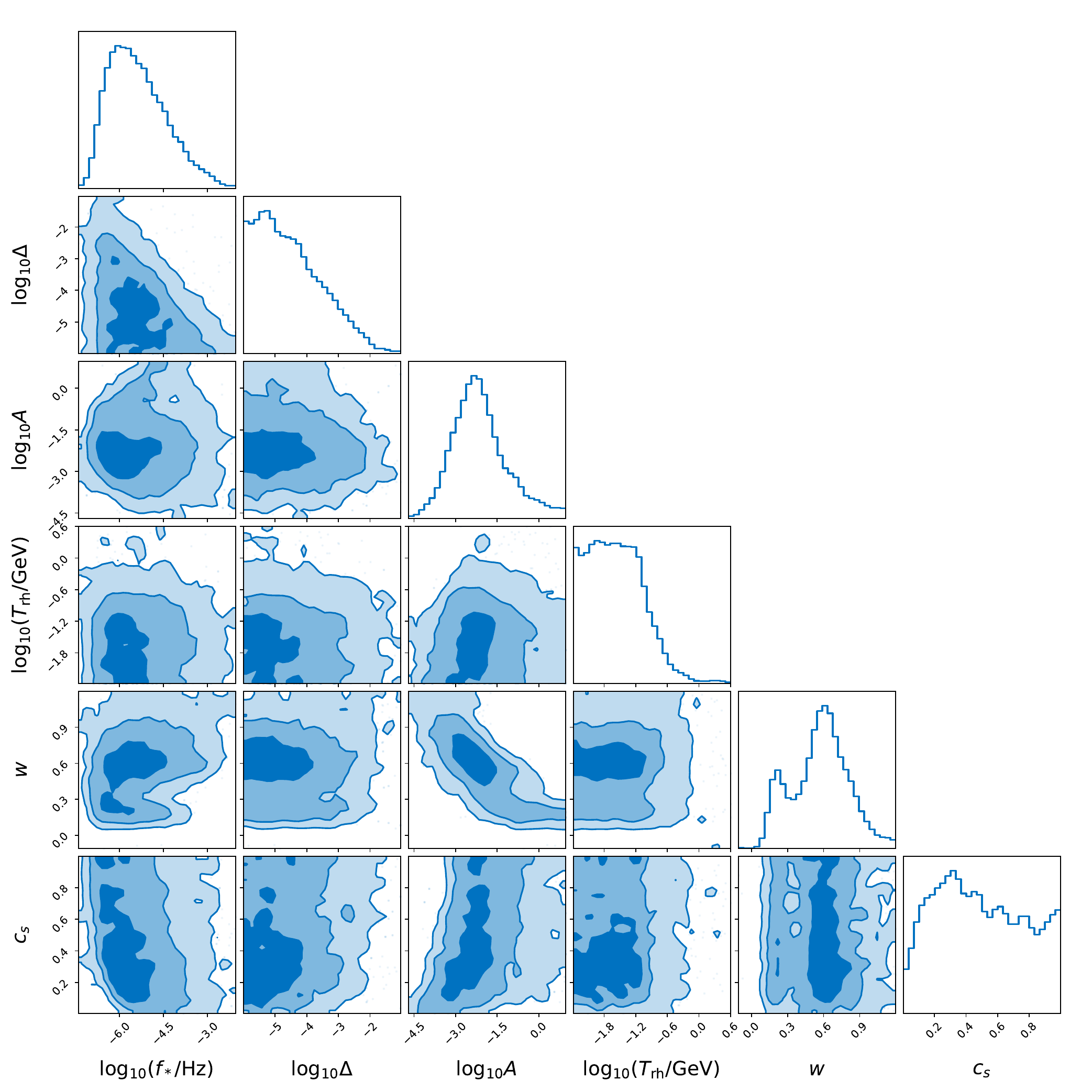}
	\caption{\label{posts_w}One and two-dimensional marginalized posteriors for the model parameters of $M_1$ obtained from the combined NANOGrav 15-yr data set, PPTA DR3, and EPTA DR2. The contours in the two-dimensional plot correspond to the $1 \sigma$, $2 \sigma$, and $3 \sigma$ credible regions, respectively.}
\end{figure}

\begin{figure}[tbp!]
	\centering
 \includegraphics[width=\textwidth]{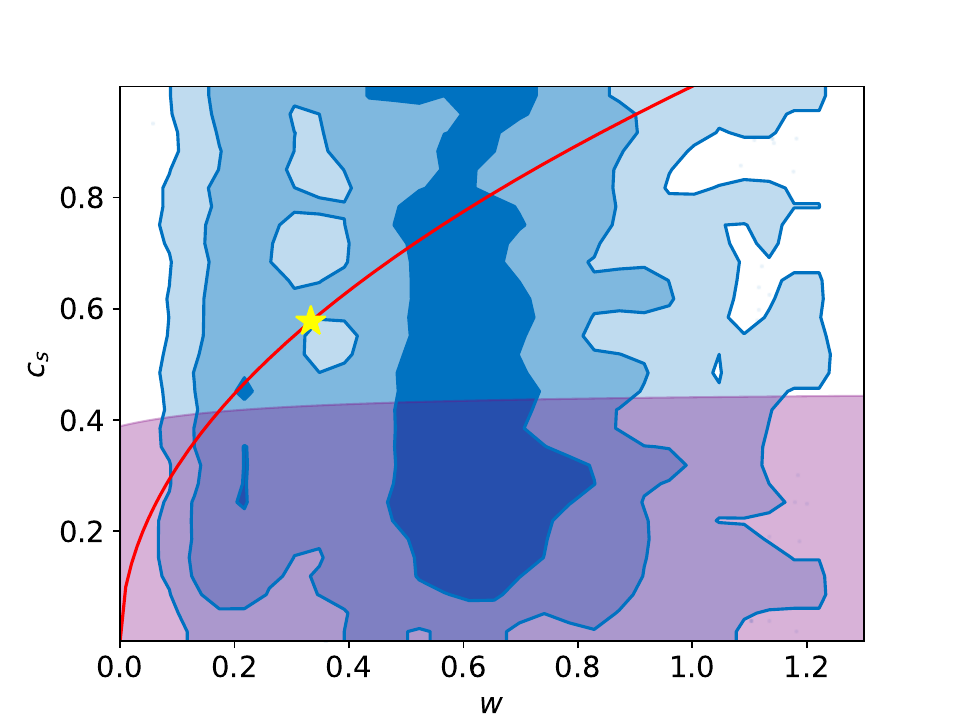}
	\caption{\label{post_w_cs} The two-dimensional posteriors for the $w$ and $c_s$ parameters shown in blue contours. The purple region represents the excluded parameter space where $\fpbh > 1$. The red curve corresponds to $c_s^2=w$, while the yellow star denotes $c_s^2=w=1/3$. Other model parameters are fixed as $\log_{10} A = -2.23$, $\log_{10} (f_*/\mathrm{Hz}) = -5.52$, $\log_{10} \Delta=-4.67$, and $\log_{10} (T_{\mathrm{rh}}/\mathrm{Gev}) = -1.6$.}
\end{figure}

We present the posterior distributions for the model parameters of $M_1$ in \Fig{posts_w}. Our analysis confirms that the PTA data favour a narrow curvature power spectrum with $\Delta \lesssim 0.003$ at the $95\%$ confidence level when considering the effect of both the EoS and sound of speed~\cite{Liu:2023pau}. This outcome validates our assumption of a narrow peak in the spectrum.
Additionally, the peak frequency is tightly constrained to $\log_{10} (f_*/\mathrm{Hz})=-5.52^{+2.00}_{-1.15}$. Furthermore, our analysis places an upper limit on the reheating temperature, with $T_\mathrm{rh} \lesssim 0.2\,\mathrm{GeV}$. Notably, to meet the constraints imposed by BBN, we require $T_\mathrm{rh} \gtrsim 4\,\mathrm{MeV}$.
Moreover, the amplitude is constrained to be $\log_{10} A = -2.23^{+2.07}_{-1.30}$, and the EoS parameter is determined to be $w = 0.59^{+0.36}_{-0.40}$. While we allow the prior for $w$ to span the range of $[-1/3, 1.2]$, our results confidently exclude negative values of the EoS parameter at $95\%$ confidence level. Instead, the data prefers an EoS parameter with $w < 1$, although it is noteworthy that $w=1/3$ remains consistent with the PTA data.
Additionally, our analysis confirms a previously known method to suppress PBH formation by increasing the sound speed~\cite{Balaji:2023ehk} as illustrated from \Fig{post_w_cs}, where the purple region indicates the parameter space excluded due to the requirement $\fpbh \leq 1$.

In \Table{tab:BF}, we summarize the Bayes factors, comparing various models to the power-law spectrum model where $\Omega_{\mathrm{GW}} \propto A^2 f^{-\gamma}$. Specifically, we consider seven models: the fiducial model $M_0$ with $c_s^2=w=1/3$, model $M_1$ with both $c_s$ and $w$ as free parameters, model $M_2$ with $c_s$ and $w$ fixed to their best-fit values ($c_s^2 =0.47$, and $w=0.59$), model $M_3$ with $c_s^2=1$ and $w$ as a free parameter, model $M_4$ with both $c_s$ and $w$ as free parameters, but requiring $c_s^2 = w$, model $M_5$ with $c_s$ as a free parameter and $w=1/3$, and the pressure-less fluid domination model model $M_6$ with $c_s$ as a free parameter and $w=0$. 
Based on the Bayes factors presented in \Table{tab:BF}, it is evident that, as expected, model $M_2$ offers the best description of the PTA data. 
Here, we want to remind readers that $M_2$, whose parameter values are fixed to the best-fit values of $M_1$, is a toy model and is included as a reference.
Distinguishing between models $M_0$ through $M_5$ proves to be a challenging task due to their similar Bayes factors.
Notably, we find that the pressure-less fluid domination model $M_6$, which has a Bayes factor of $0.0055$, is confidently ruled out.

\section{\label{Con}Summary and discussion}

While observations of the CMB and large-scale structure have provided increasingly precise measurements on the largest scales of the Universe, our understanding of small scales remains limited, except for the insights provided by PBHs. PTAs, on the other hand, offer invaluable means to investigate the characteristics of the early Universe through the SIGWs. Assuming that the stochastic signal detected by PTA collaborations originates from SIGWs, we have jointly employed the NANOGrav 15-year data set, PPTA DR3, and EPTA DR2 to constrain the conditions of the early Universe. In this paper, we have simultaneously imposed constraints on the sound speed and EoS parameter as well as the reheating temperature, yielding $w = 0.59^{+0.36}_{-0.40}$ (median + $90\%$ credible interval), and $T_\mathrm{rh}\lesssim 0.2\,\mathrm{GeV}$ at the $95\%$ credible interval for a lognormal power spectrum of the curvature perturbation. It is worth noting that although our analysis focused on the lognormal power spectrum of curvature perturbations, the methodology and framework presented in this work can be easily extended to other types of power spectra.

To further evaluate the plausibility of different models, we have calculated Bayes factors by comparing them with the power-law spectrum model where $\Omega_{\mathrm{GW}} \propto A^2 f^{-\gamma}$. Interestingly, our analysis reveals that the pressure-less fluid domination model is ruled out because $w=0$ corresponds to a flat SIGW spectrum that is obviously inconsistent with the PTA data. Moreover, the data also provides support for radiation domination ($w=1/3$) within a $90\%$ credible interval. The results of our study demonstrate the potential of SIGWs as a powerful tool for exploring the nature of the early Universe.

It is worth noting that the precision of our analysis hinges on the reliability of the noise modeling implemented by each PTA collaboration. The PPTA, EPTA, and NANOGrav have each embraced unique approaches to noise modeling for their pulsars. Particularly, the PPTA~\cite{Reardon:2023zen} has undertaken an extensive noise analysis, uncovering the significant impact of noise modeling intricacies on spectral characterization. Therefore, while our analysis is built upon the available PTA data releases, we acknowledge the potential for alterations in our results should the noise models undergo updates.

\section*{Acknowledgments}
We are grateful to Qing-Guo Huang for the useful discussions.
LL is supported by the National Natural Science Foundation of China (Grant No.~12247112 and No.~12247176) and the China Postdoctoral Science Foundation Fellowship No. 2023M730300.
ZCC is supported by the National Natural Science Foundation of China (Grant No.~12247176 and No.~12247112) and the China Postdoctoral Science Foundation Fellowship No. 2022M710429.
\bibliographystyle{JHEP}
\bibliography{ref}

\end{document}